\documentstyle[amssymb,preprint,aps]{revtex}

\begin{document}
\draft
\author{M. Jaime\thanks{%
Present address Los Alamos National Laboratory, MSK764, Los Alamos, NM 87545}%
, P. Lin, S. H. Chun, M. B. Salamon P. Dorsey$^{\S }$ and M. Rubinstein$^{\S
}$}
\address{Department of Physics and Materials Research Laboratory\\
University of Illinois at Urbana-Champaign\\
1110 W. Green Street, Urbana, IL 61801\\
\S\ U.S. Naval Research Laboratory, Washington D.C. 20375-5000}
\title{Coexistence of localized and itinerant carriers near T$_{C}$ in
calcium-doped manganites}
\date{
\today%
. revised version}
\maketitle

\begin{abstract}
We explore the possibility that polaronic distortions in the paramagnetic
phase of La$_{0.67}$Ca$_{0.33}$MnO$_{3}$ manganites persist in the
ferromagnetic phase by considering the observed electrical resistivity to
arise from coexisting field- and temperature-dependent polaronic and
band-electron fractions. We use an effective medium approach to compute the
total resistivity of the two-component system, and find that a limit with
low percolation treshold explains the data rather well. To test the validity
of this model, we apply it to the thermoelectric coefficient. We propose a
plausible mean-field model that reproduces the essential features of a
microscopic model and provides a comparison with the experimental mixing
fraction, as well as the magnetization and magnetic susceptibility.
\end{abstract}


\tightenlines

There is a growing consensus that the so-called colossal magnetoresistivity
(CMR) of La$_{1-x}$ A$_x$MnO$_3,$ where A is a divalent substituent, is
larger when both {\it double exchange }and{\it \ strong coupling to local
lattice deformations }are involved.\cite{millis,roder,exp,jaime3} In the
double exchange model,\cite{zener,koh} electrons can hop from the singly
occupied e$_{2g}$ orbital of Mn$^{3+}$ ions to empty e$_{2g}$ orbitals of
neighboring Mn$^{4+}$ ions. Strong Hund's-rule coupling enhances the hopping
matrix element when the S = 3/2 $t_{2g}$ cores of the neighboring sites are
aligned, thereby favoring ferromagnetism and an increased bandwidth.
However, as Millis and coworkers \cite{millis,millis2} have emphasized, the
Jahn-Teller effect in Mn$^{3+},$ if strong enough, can lead to polaron
formation and the possibility of self-trapping. The effective Jahn-Teller
coupling constant $\lambda _{eff}$, in this picture, must be determined
self-consistently, both because it depends inversely on the bandwidth and
because the effective transition temperature increases with decreasing $%
\lambda _{eff}.$ If $\lambda _{eff}$ is larger than a critical value $%
\lambda _c$, the system consists of polarons in the paramagnetic phase. As
the temperature is lowered to the Curie temperature T$_C,$ the onset of
ferromagnetism increases the effective bandwidth, which reduces $\lambda
_{eff},$ thereby increasing the effective transition temperature. As a
result, the polarons may dissolve into band electrons if $\lambda _{eff}$
drops below $\lambda _c$ and the material reverts to a half-metallic, double
exchange ferromagnet at low temperatures.

In the Millis {\it et al.} model, the tendency toward polaron formation is
monitored by a local {\em displacement} coordinate $r$, which is zero for $%
\lambda _{eff}<\lambda _c,$ and grows continuously as $\lambda _{eff}$
increases beyond that value. However, polarons are typically \cite{emin}
bimodal--large or small--so that we should consider $r$ to be a measure of
the relative proportion of large polarons (band electrons for which $%
r\approx 0$) and small polarons (for which $r$ is an atomic scale length).
Indeed, there is growing experimental evidence\cite
{louca,yoon,booth,billinge} that polaronic distortions, evident in the
paramagnetic state, persist over some temperature range in the ferromagnetic
phase. This paper explores that possibility, in the effective medium
approximation, by considering the observed electrical resistivity to arise
from a field- and temperature-dependent concentration $c(H,T)$ of metallic
regions within a semiconductive, polaronic background having an activated
electrical conductivity. \cite{landauer} We test the validity of this model
by applying it to the thermoelectric coefficient, also within the effective
medium approximation, and extract $c(H,T)$ for stripe-like domains. We
propose a mean-field model, in which $c(H,T)$ is a secondary order
parameter, that reproduces the qualitative features of the experimental data.

The La$_{2/3}$Ca$_{1/3}$MnO$_3$ film samples used in this study were
prepared by pulsed laser deposition onto LaAlO$_3$ substrates to a thickness
of 0.6 $\mu $m$.$ As described previously \cite{jaime2}, they were annealed
at 1000 ${{}^{\circ }}$C for 48 hr. in flowing oxygen. Measurements were
carried out in a 7 T Quantum Design Magnetic Property Measurement System
with and without an oven option provided by the manufacturer. A modified
sample rod brought electrical leads and type-E thermocouples to the sample
stage. A bifilar coil of 12 $\mu $m Pt wire was calibrated to serve both as
a thermometer and to provide a small heat input for the thermopower
measurements. Measurements in fields up to 7 T could be carried out over the
temperature range $4$ K $\leq $ $T\leq 500$ K. Following the transport
measurements, magnetization data $M(H,T)$ were acquired up to $380$ K by
conventional methods.

Figure 1 shows the magnetization and resistivity data over the temperature
range of interest. The data below $200$ K exhibit essentially field
independent, metallic behavior, and are well fit by a power law,

\begin{equation}
\rho _{lt}(T)=[0.22+2\times 10^{-5}\text{ K}^{-2}T^{2}+1.2\times 10^{-12}%
\text{ K}^{-5}T^{5}]\text{ m}\Omega \text{ cm.}  \label{eq1}
\end{equation}

Above 260 K, the resistivity is exponential and again independent of field,
given \cite{jaime} by the form expected for the adiabatic hopping of small
polarons,

\begin{equation}
\rho _{ht}(T)=(1.4\mu \Omega \text{ cm K}^{-1})T\exp (1276\text{ K}/T).
\label{eq2}
\end{equation}
Both limits are displayed in Figure 1 as dotted lines. Our assumption is
that these represent the resistivity of band electrons and polarons,
respectively, and that the transition region can be represented by the
conductivity of an effective medium \cite{landauer} characterized by a
mixing factor $c(H,T)$ which we envisage to be the fraction of the carriers
that are in the metallic state. We assume the effective resistivity $\rho
(H,T)$ to satisfy the general expression for ellipsoidal metallic regions
(resistivity $\rho _{lt}(T)$) embedded in a semiconducting matrix
(resistivity $\rho _{ht}(T)$) \cite{ls}, 
\begin{equation}
\frac{c\left( \rho -\rho _{lt}\right) }{3}\left[ \frac{1}{\rho
_{lt}+g_{\Vert }(\rho -\rho _{lt})}+\frac{2}{\rho _{lt}+g_{\bot }(\rho -\rho
_{lt})}\right] +\frac{3(1-c)(\rho -\rho _{ht})}{2\rho _{ht}+\rho }=0.
\label{eq3}
\end{equation}
Here, $g_{\bot }\simeq $ 1/2, $g_{\Vert }=(b^{2}\rho _{ht}/a^{2}\rho )\ln
[1+(a\rho /b\rho _{ht})]$ and $a,$ $b$ are the (major, minor) axes of the
prolate ellipsoids. As a first approximation, we set $c(0,T)=M(0,T)/M_{sat}$%
, using the data of Fig. 1{\it a}. The chain curves in Fig. 1b show that the
magnetization roughly reproduces the shape of the resistivity curves at zero
field. That the metallic resistivity switches on quickly as the system
orders indicates a low percolation threshold and a large aspect ratio ($a>>b$%
). This suggests that the metallic regions appear as stripe domains, rather
than as compact clusters. The magnetization does not approximate, at all,
the relative concentrations in applied fields, as seen in the solid line in
Fig. 1b. As an alternative, we solve Eq. (3) for $c(H,T)$ using the
experimental resistivity $\rho _{\exp }(H,T)$ , with the result shown in
Fig. 2a for $a/b$ = 50.

As a second test of this approach, we consider the Seebeck coefficient $%
S(H,T)$, measured over the same temperature range, and plotted in Fig. 2{\it %
b}. We fit the low temperature thermopower arbitrarily to a power law 
\begin{equation}
S_{lt}(T)=[(0.051\text{ K}^{-1})T-(1.3\times 10^{-4}\text{ K}%
^{-2})T^2-(3.2\times 10^{-7}\text{ K}^{-3})T^3]\text{ }\mu \text{V/K,}
\label{eq4}
\end{equation}
and the high temperature data \cite{jaime} to the form expected for small
polarons, 
\begin{equation}
S_{ht}(T)=[(9730\text{ K})T^{-1}-29]\text{ }\mu \text{V/K.}  \label{eq5}
\end{equation}
Low and high temperature limits are displayed as dotted lines in Figure 2%
{\it b}. A remarkable feature of the transport properties of composites is
that knowing any two effective transport properties and those of the
constituents determines the third. \cite{straley} A closed expression for
the Seebeck coefficient has been derived by Bergman and Stroud \cite{bs}
which, by recognizing that the effective conductivities match the high and
low temperature values in the respective limits, we can write in the
following form, 
\begin{equation}
S(H,T)=\frac{\left( S_{lt}-S_{ht}\right) \left( \kappa _{lt}-\kappa
_{ht}\right) }{\left( \kappa _{ht}\text{ }\rho _{ht}-\kappa _{lt}\text{ }%
\rho _{lt}\right) }\frac{\rho _{lt}\text{ }\rho _{ht}}{\left( \rho
_{ht}-\rho _{lt}\right) }+\frac{\left( S_{lt}\text{ }\rho _{ht}-S_{ht\text{ }%
}\rho _{lt}\right) }{\left( \rho _{ht}-\rho _{lt}\right) }+\frac{\left(
S_{ht}-S_{lt}\right) \kappa (H,T)\text{ }\rho (H,T)}{\left( \kappa _{ht}%
\text{ }\rho _{ht}-\kappa _{lt}\text{ }\rho _{lt}\right) }.  \label{eq6}
\end{equation}
Here, $\kappa _{lt}$, $\kappa _{ht},$ and $\kappa (H,T)$ are the thermal
conductivities of low and high temperature phases and the effective medium,
respectively. While small changes in $\kappa (H,T)$, of order 20\% in the
transition region, have been reported \cite{ramirez}, the thermal
conductivity of this thin-film sample is dominated by the substrate. \ All
three thermal conductivities are, therefore, equal at each temperature-field
point, enabling us to simplify the expression to, 
\begin{equation}
S(H,T)=\frac 1{\rho _{ht}-\rho _{lt}}\left[ \rho _{ht}S_{lt}-\rho
_{lt}S_{ht}+\rho _{\exp }(H,T)(S_{ht}-S_{lt})\right] .  \label{eq7}
\end{equation}
The results are shown as solid lines in Fig. 2{\it b}; the agreement is
excellent, with no adjustable parameters. Measurements using single-crystal
samples are underway to explore the effect of $\kappa (H,T)$ on the analysis.

The precise shape of the $c(H,T)$ curves depends on the form of the
effective medium model chosen, being sharper the lower the percolation
threshold. Nonetheless, it is clear that the metallic fraction increases
rapidly below the transition temperature, but does not broaden in applied
field to the same extent as the magnetization. It is useful, therefore, to
explore under what circumstances this can occur. The essential feature of
the Millis {\it et al.} model is that the effective Jahn-Teller coupling
constant $\lambda _{eff}$ is very near its critical value in the
paramagnetic phase. There are two phase transitions to consider a
zero-temperature polaronic transition at a critical coupling constant $%
\lambda _c$ that would occur in the absence of Hund's rule coupling, and a
double-exchange ferromagnetic transition at $T_C$ for $\lambda _{eff}=0$.
The dependence of the bandwidth on magnetic order in the double exchange
picture causes both $\lambda _{eff}$ and $T_C$ to depend on the temperature
through the magnetization. We propose here a simplified mean-field model
that reproduces the essential features of the microscopic calculation to
demonstrate that $c(H,T)$ is a secondary order parameter, driven by the
primary order parameter, namely the magnetization, and that it resembles the
experimentally determined curves. We assume that the ferromagnetic
free-energy functional is of conventional form 
\begin{equation}
F_{mag}=\frac 12(T/T_C-1)m^2+\frac 14bm^4-mh,  \label{eq8}
\end{equation}
written in units of \cite{mattis} $3Sk_BT_C/(S+1)=1.94k_BT_C$ for $%
S=2(1-x)+3x/2=1.83$ and $x=1/3.$ Further, we have $h=g\mu
_B(S+1)H/3k_BT_C=H/2360$ kOe. The secondary parameter $c(H,T)$ is asumed to
be driven by the difference $\lambda _{eff}-\lambda _c.$ We approximate the
dependence of $\lambda _{eff}$ on the magnetization by writing $\lambda
_{eff}-\lambda _c\varpropto \alpha -\gamma m^2+...,$ where $\alpha $ is
small and positive. The electronic free energy can then be written, in the
same dimensionless units as Eq. (\ref{eq8}), as 
\begin{equation}
F_{el}=\frac 12(\alpha -\gamma m^2)c^2+\frac 14\beta c^4.  \label{eq9}
\end{equation}
Minimizing the total free energy $F=F_{mag}+F_{el}$, we obtain 
\begin{equation}
\frac{\partial F}{\partial m}=(T/T_C-1-\gamma c^2)m+bm^3-h=0  \label{eq10}
\end{equation}
\begin{equation}
\frac{\partial F}{\partial c}=(\alpha -\gamma m^2)c+\beta c^3=0  \label{eq11}
\end{equation}
From Eq. (\ref{eq11}) it is obvious that the concentration of metallic
electrons is zero until the magnetization reaches the value $m=\sqrt{\alpha
/\gamma },$ beyond which point $c$ increases. In the limit $\alpha
\rightarrow 0,$ $c$ is proportional to $m$, with the result that $%
b\rightarrow b-\gamma ^2/\beta ,$ signalling a tendency for the system to
approach a tricritical point and first order transitions as the coupling
constant is increased. Note that the existence of a non-zero concentration $%
\overline{c}$ can be considered to increase the critical point to $(1+\gamma 
\overline{c}^2)T_C,$ causing the magnetization to increase more rapidly than
would be the case without coupling to the metallic electron concentration.

To proceed, we recognize that Eq.(\ref{eq10}) is the expansion of a
Brillouin function, 
\begin{equation}
m=B_S\left( \frac{3ST_C}{(S+1)T}[(1+\gamma c^2)m+h)\right) ,  \label{eq12}
\end{equation}
while the equation for $c$ can be assumed to be a small-$c$ expansion of 
\begin{equation}
c=\tanh \left[ (1-\alpha +\gamma m^2)c\right] .  \label{eq13}
\end{equation}

In Fig. 3 we show the simultaneous solutions of Eqs. (\ref{eq12} \& \ref
{eq13}) for $\alpha =0.02$ and $\gamma =0.3$ at $H=0,$ $24$ kOe, and $48$
kOe. Application of the magnetic field increases the temperature at which $c$
becomes non-zero by 7\% or 20 K, consistent with the experimental data in
Fig. 3 and, unlike the magnetization, does produce a high-temperature tail
in $c$. As no thermal factors are included in the definition of $c,$ the
concentration of free carriers does not approach unity, and therefore
differs slightly from the experimental $c(H,T)$ obtained from Eq. (\ref{eq3}%
). The abrupt appearance of band electrons in this model produces a kink in
the zero field magnetization curve at the onset temperature, seen as a
deviation from the $H=0$, $\gamma =0$ curve. The inset to Fig. 4 shows the
inverse of magnetization data in an applied field of 1 kG taken on a single
crystal of La$_{0.7}$Ca$_{0.3}$MnO$_3$, \cite{chun} along with H/m for the
24 kG data of Fig. 3. The experimental kink moves to higher temperature as
the field is increased, as predicted by the model, but occurs further from
the actual transition than the model predicts. The transition temperature
for this sample, obtained by a scaling analysis of the data below the kink
temperature, is 216 K. Clearly, the mean-field model proposed here cannot
capture the precursive behavior due to critical fluctuations, important in
the actual experiment, that would move the kink to higher temperatures.

As noted above, a number of other experimental probes have suggested the
coexistence of localized and delocalized d-holes in the ferromagnetic state.
Booth et al. \cite{booth} have defined the density of delocalized holes $%
n_{dh}$ analogously to our Eq.\ (\ref{eq3}) from the width of the Mn-O EXAFS
peak. They suggest that $n_{dh}\varpropto \exp (3.5m),$ which leaves a
finite density of delocalized carriers above $T_C$ and a nearly linear
increase in $n_{dh}$ below. The fits for the mean-square width of the Mn-O
bond length distribution, however, look remarkably like the magnetization
itself. Louca, et al. \cite{louca}, in a Sr-doped sample with comparable $%
T_C $, report a gradual change in the number of nearest Mn-O neighbors as
the temperature is reduced below the critical point. They interpret their
data in terms of a gradual transition from single-site to triple-site
polarons, rather than to free carriers as favored here. Similar neutron data
by Billinge, et al. \cite{billinge} show that the O-O bond-length
distribution begins to narrow below $T_C$ in a manner similar to our $%
c(0,T). $ Recent Raman results \cite{yoon} find two contributions in the
Raman intensity a diffusive component associated with small polarons and a
continuum contribution from free carriers. The relative intensities of these
components have temperature dependences that are very similar to $1-c(0,T)$
and $c(0,T),$ respectively.

That the field and temperature dependence of the thermopower can be deduced
directly from resistivities and field-independent component Seebeck
coefficients lends strong support to a picture in which conducting regions
arise and percolate in the vicinity of the phase transition. The rapid
decrease in resistance requires a low percolation threshold which in turn,
points to stick-like, or stripe-like domains. \cite{cheong} Our simple
mean-field model ignores a number of features that should be included in a
complete treatment. In particular, we have excluded a term $m^2c$ because it
leads to a first-order transition for all values of the parameters; we
cannot rule it out on symmetry grounds. Similarly, there should be a mixing
entropy in the electronic free energy which, at sufficiently high
temperatures, will lead to thermal dissociation of the polarons. Finally, we
have not included gradient terms and therefore ignore inhomogeneous thermal
fluctuations that are certain to be significant in a system such as this
where there are competing order parameters. Nonetheless, our
phenomenological approach provides a qualitative understanding of the field
and temperature dependence of the transport properties while correctly
predicting the existence of kinks in the magnetization curves.

This work was supported in part by DOE Grant No. DEFG0291ER45439 through the
Illinois Materials Research Laboratory. MJ acknowledges support from U.S.
Department of Energy at Los Alamos National Laboratory, NM.

\begin{figure}
\caption{ (a) Magnetization data for this sample. (b) Resisitivity data as functions of field
and temperature.  The broken curves are effective field calculations at two values
of $a/b$, assuming that the low field magnetization represents the metallic concentration. 
The solid line uses the magnetization measured at 10 kOe, and $a/b = 50$. Dotted lines are 
the low and high temperature limits discussed in the text.}
\label{ fig1}
\end{figure}%

\begin{figure}
\caption{(a) The mixing factor $c(H,T)$ extracted from the resisitivity using the effective
medium approximation with $a/b=50$. (b) Seebeck coefficient data and results of a 
computation using the measured resistivity $\rho_{exp} (H,T)$ as described in the text. Dotted 
lines show the low and high temperature fits.}
\label{ fig2}
\end{figure}%

\begin{figure}
\caption{(a) The mixing factor $c(H,T)$ calculated in the mean-field model with $\alpha
=0.02$ and $\gamma = 0.3$. (b) The magnetization calculated with the same 
parameters. The dotted line shows the non-interactive case for comparison. }
\label{fig3 }
\end{figure}%

\begin{figure}
\caption{Inverse magnetic susceptibility $H/m$ $vs$ temperature near the M-I transition 
for $H=24$ kOe. The dashed line is the zero field extrapolated behavior. The appearance 
of free carriers induce the rise of the effective T$_{C}$, leading to a kink in the 
susceptibility.  Inset Inverse susceptibility of a single crystal sample of 
La$_{0.7}$Ca$_{0.3}$MnO$_3$,  measured in a field of 1 kG. }
\label{fig5 }
\end{figure}%


\begin{references}
\bibitem{millis}  A. J. Millis, P.B. Littlewood, and B. I. Shraiman, Phys.
Rev.Lett.{\bf 74}, 5144 (1995).

\bibitem{roder}  H. R\"{o}der, J. Zang, and A. R. Bishop, Phys. Rev. Lett. 
{\bf 76}, 1356 (1996).

\bibitem{exp}  J. Tanaka {\it et al.}, Jou. Phys. Soc. Japan {\bf 51} 1236
(1982), H.Y. Hwang {\it et al.}, Phys. Rev. Lett. {\bf 75}, 914 (1995), M.F.
Hundley {\it et al.}, Appl. Phys. Lett. {\bf 67}, 860 (1995), M.R. Ibarra 
{\it et al.}, Phys. Rev. Lett.{\bf \ 75}, 3541 (1995), J.J. Neumeier {\it et
al.}, Phys. Rev. B{\bf 52}, R7006 (1995), J.-H. Park {\it et al.}, Phys.
Rev. Lett.{\bf \ 76}, 4215 (1996), P. Dai {\it et al.}, Phys. Rev. B{\bf 54}%
, R3694 (1996), D.N. Argyriou {\it et al.}, Phys. Rev. Lett. {\bf 76}, 3826
(1996), G. Zhao {\it et al.}, Nature {\bf 381}, 676 (1996). R.H. Heffner{\it %
\ et al.}, Phys. Rev. Lett. {\bf 77}, 1869 (1996), etc.

\bibitem{jaime3}  M. Jaime, H.T. Hardner, M.B. Salamon, M. Rubinstein, P.
Dorsey, and D. Emin, Phys. Rev. Lett. {\bf 78} 951 (1997). Also M. Jaime 
{\it et al}., J. Appl. Phys. {\bf 81}, (1997).

\bibitem{zener}  C. Zener, Phys. Rev. {\bf 82, }403 (1951); P. W. Anderson
and H. Hasegawa, Phys. Rev. {\bf 100}, 675 (1955).

\bibitem{koh}  K. Kubo and N. Ohata, J. Phys. Soc. Jpn., {\bf 33}, 21 (1972).

\bibitem{millis2}  A. J. Millis, R. Mueller, and B. I. Shraiman, Phys. Rev. 
{\bf 54}, 5389 (1996), {\it ibid. }{\bf 54}, 5405 (1996).

\bibitem{emin}  D. Emin and T. Holstein, Phys. Rev. Lett. {\bf 36}, 1492
(1976).

\bibitem{louca}  D. Louca, T. Egami, E.L. Brosha, H. R\"{o}der, and A.R.
Bishop, Phys. Rev. B {\bf 56}, R8475 (1997).

\bibitem{yoon}  S. Yoon, H.L. Liu, G. Schollerer, S.L. Cooper, P.D. Han,
D.A. Payne, S.-W. Cheong, and Z. Fisk, Phys. Rev. B {\bf 58}, 2795 (1998).

\bibitem{booth}  C. H. Booth, F. Bridges, G.H. Kwei, J.M. Lawrence, A.L.
Cornelius, and J.J. Neumeier, Phys. Rev. Lett. {\bf 80}, 853 (1998). Similar
results are discussed by A. Lanzara {\it et al.}, Phys. Rev. Lett. (to be
published).

\bibitem{billinge}  S. J. L. Billinge, R.G. DiFrancesco, G.H. Kwei, J.J.
Neumeier, and J.D. Thompson, Phys. Rev. Lett. {\bf 77}, 715 (1996).

\bibitem{landauer}  R. Landauer in {\it Electrical Transport and Optical
Properties of Inhomogeneous Media}, edited by J. C. Garland and D. B. Tanner
(American Institute of Physics, New York, 1978) pg. 2.

\bibitem{jaime2}  M. Jaime, M. B. Salamon, K. Pettit, M. Rubinstein, R.E.
Treece, J.S. Horwitz, and D.B. Chrisey, J. Appl. Phys. {\bf 68}, 1576 (1996).

\bibitem{jaime}  M. Jaime, M.B. Salamon, M. Rubinstein, R.E. Trece, J.S.
Horwitz, and D.B. Chrisey, Phys. Rev. B {\bf 54}, 11 914 (1996)

\bibitem{ls}  A. N. Lagarkov and A. K. Sarychev, Phys. Rev. B {\bf 52}, 6318
(1996).

\bibitem{straley}  J. P. Straley, J. Phys. D {\bf 14}, 2101 (1981).

\bibitem{bs}  D. Bergman and D. Stroud, in {\it Solid State Physics, vol. 46}%
, edited by H. Ehrenreich and D. Turnbull (Academic Press, New York, 1992)
pg. 147.

\bibitem{ramirez}  D. W. Visser, A. P. Ramirez, and M. A. Subramanian, Phys.
Rev. Lett. {\bf 78}, 3947 (1997); J.L. Cohn, J.J. Neumeier, C.P. Popoviciu,
K.J. McClellan, and T. Leventouri, Phys. Rev. B {\bf 56}, R8495 (1997).

\bibitem{mattis}  D. C. Mattis, {\it The Theory of Magnetism II},
(Springer-Verlag, Berlin, 1985) pg. 22.

\bibitem{chun}  S. H. Chun, et al. (to be published).

\bibitem{cheong}  S-W. Cheong and C. H. Chen, in {\it Colossal
Magnetoresistance and Related Properties}, edited by B. Raveau and C. N. R.
Rao (World Scientific, in press).
\end{references}
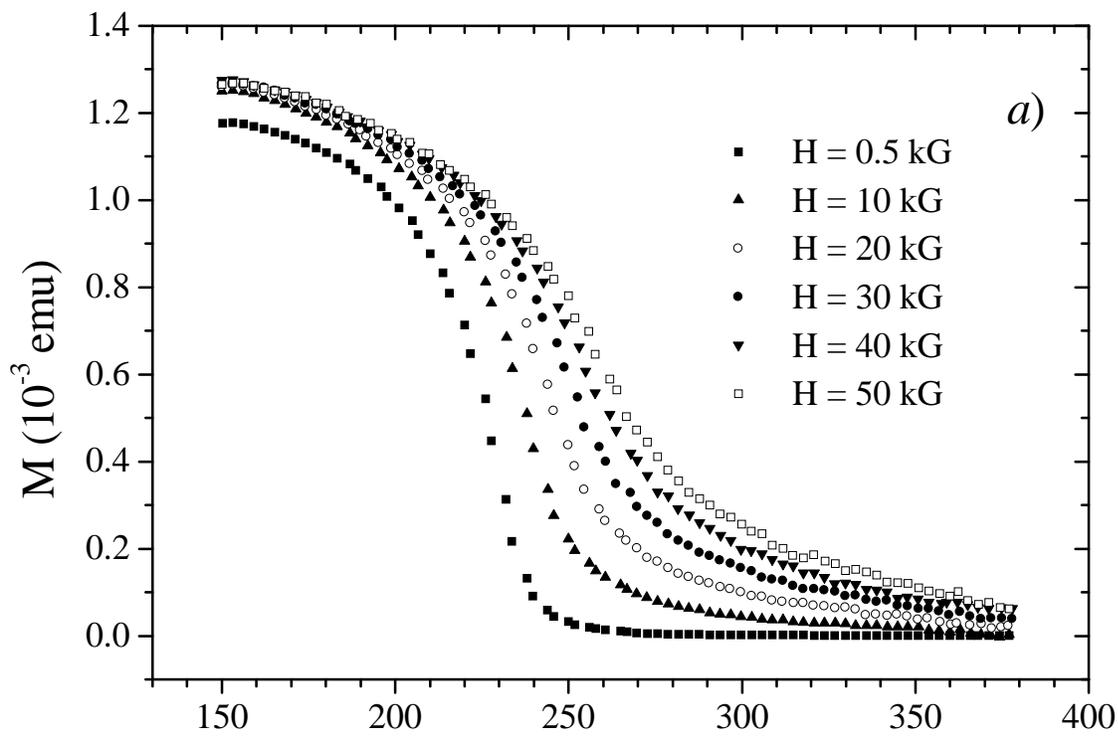
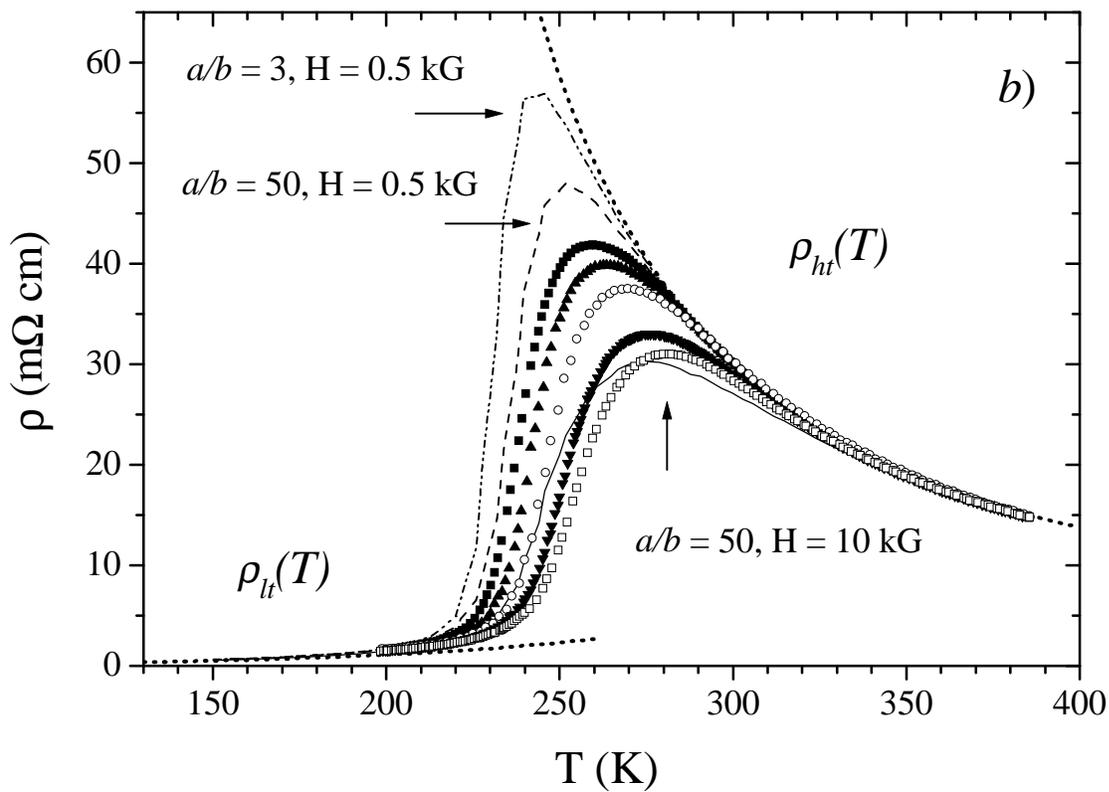
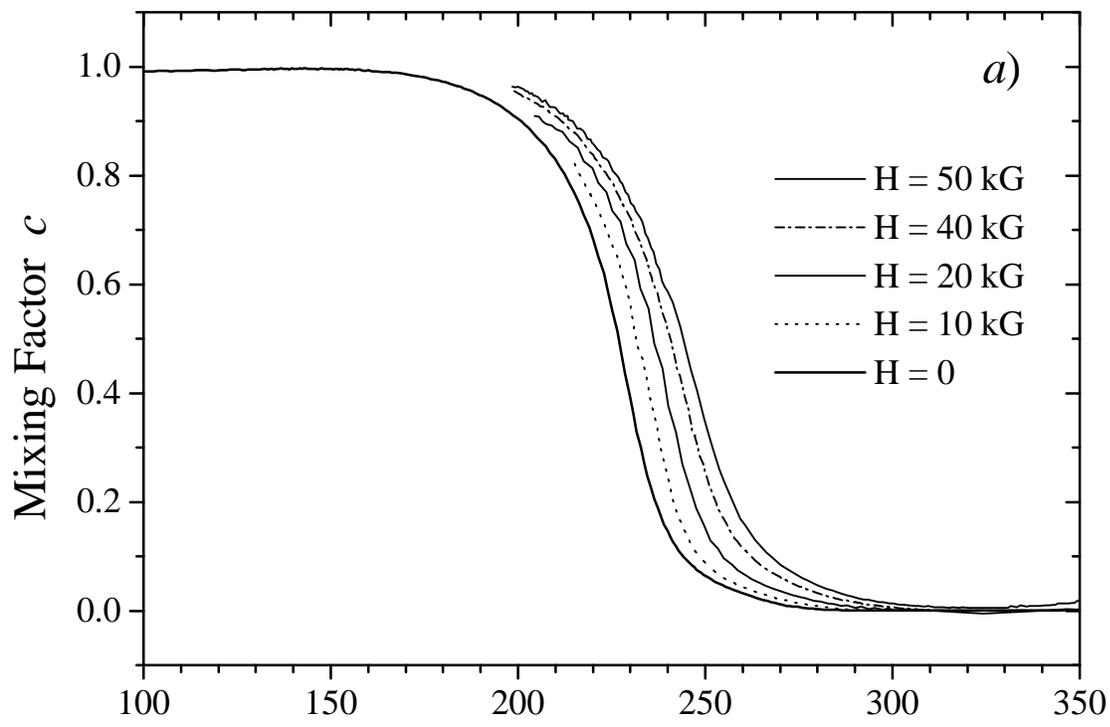
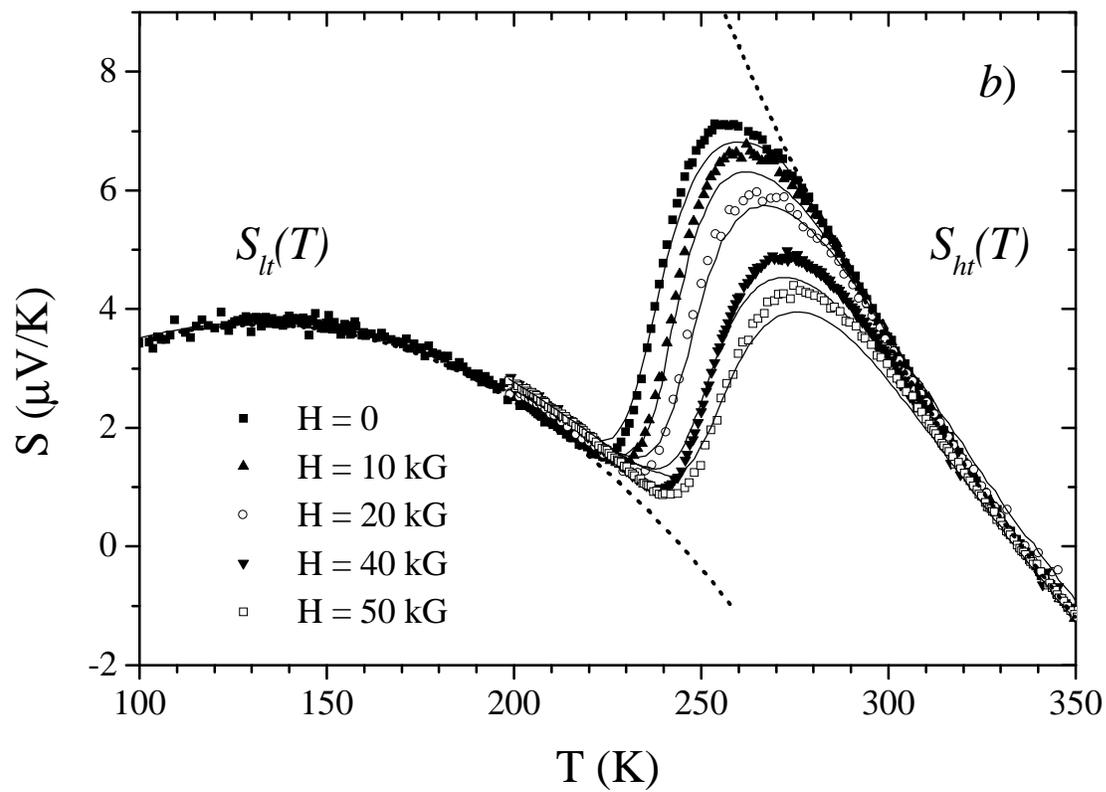
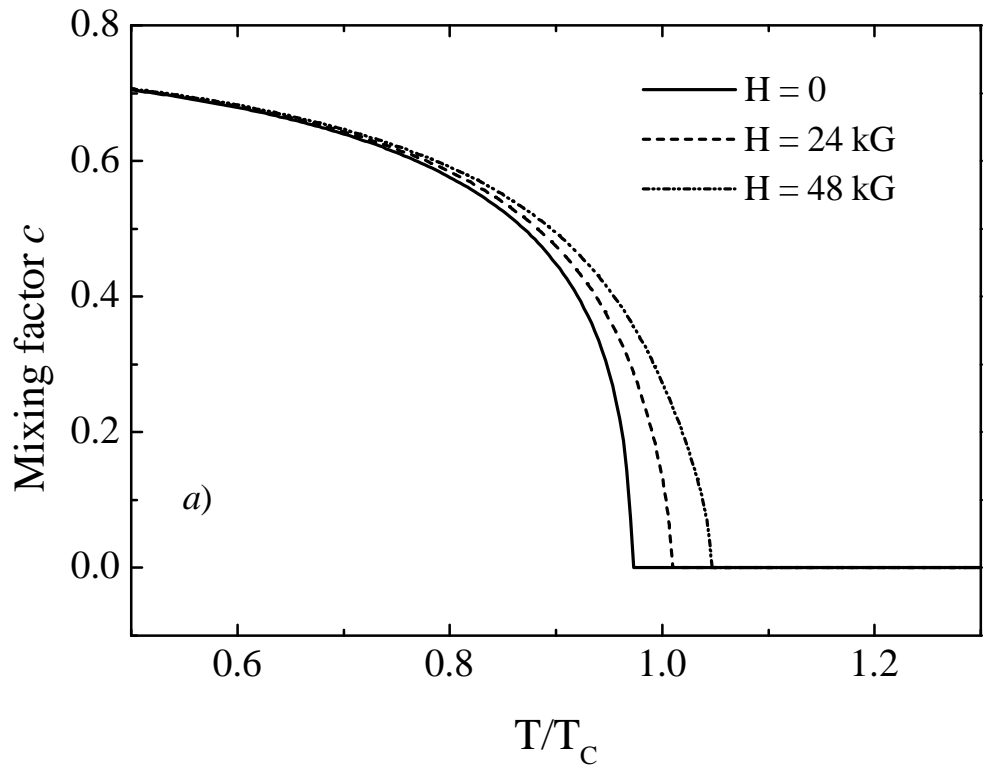
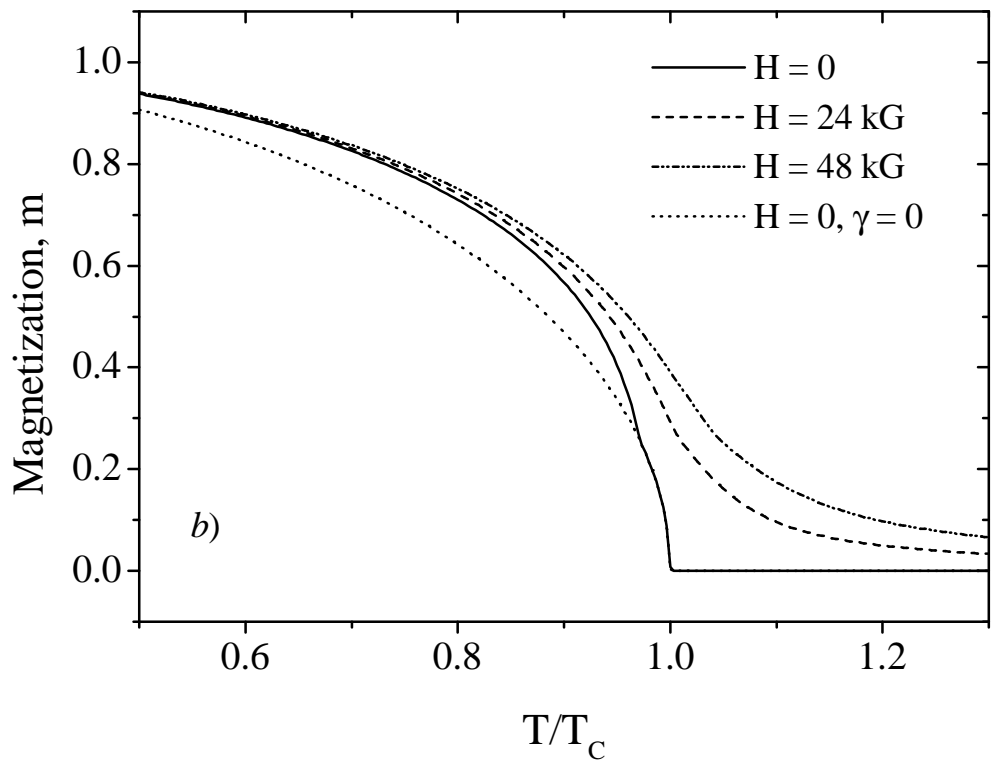
\end{document}